\documentstyle[floats,aps]{revtex}
\input epsf

\newcommand{\pbar}{\overline{p}}
\newcommand{\nbar}{\overline{n}}

\newcommand{\jy}{J/\psi}
\newcommand{\piz}{\pi^{0}}

\newcommand{\pimp}{\pi^{\mp}}
\newcommand{\kpm}{K^{\pm}}
\newcommand{\kp}{K^{+}}
\newcommand{\km}{K^{-}}
\newcommand{\ks}{K_{s}}

\newcommand{\xt}{\xi(2220)}
\newcommand{\qbar}{\overline{q}}
\newcommand{\ppair}{\pbar p}
\newcommand{\nbarn}{\nbar n}

\newcommand{\kkbar}{K\overline{K}}

\newcommand{\ksks}{\ks\ks}

\newcommand{\kpipm}{\kpm\pimp}

\newcommand{\Npair}{\overline{N}N}
\newcommand{\pbarp}{\ppair}

\newcommand{\qqbar}{q\qbar}

\newcommand{\phiphi}{\phi\phi}
\newcommand{\pptoksks}{\pbarp\rightarrow\ksks}

\newcommand{\pptonn}{\pbarp\rightarrow\nbarn}

\newcommand{\pptopizpiz}{\pbarp\rightarrow\piz\piz}
\newcommand{\fpptoksks}{\pptoksks\rightarrow 4\pi^{\pm}}

\newcommand{\pptokskr}{\pbarp\rightarrow K_{s}K^{*}}
\newcommand{\pptofourk}{\pbarp\rightarrow 4K^{\pm}}

\newcommand{\spptokskr}{\pbarp\rightarrow K_{s}K^{*}\rightarrow K_{s}K_{s}\pi^{0}}

\newcommand{\pptokskrc}{\pptokskr\rightarrow\ks\kpipm}

\newcommand{\pptophiphi}{\pbarp\rightarrow \phiphi}

\newcommand{\brjtogxi}{\rm{BR}(\jy\rightarrow\gamma\xi)}
\newcommand{\brxtopp}{\rm{BR}(\xi\rightarrow\pbarp)}
\newcommand{\brxtopbarp}{\rm{BR}(\xi\rightarrow\pbarp)}

\newcommand{\brxtoksks}{\rm{BR}(\xi\rightarrow\ksks)}

\newcommand{\degree}{\mbox{$^{\circ}$}}
\newcommand{\etal}{{\em et al.}}

\begin{document}

\title {Measurement of the $\pptoksks$ Reaction 
from 0.6 to 1.9~GeV/$c$ \\}

\author{C.~ Evangelista, A.~Palano }
\address{University of Bari and INFN, Bari, Italy}

\author{D.~Drijard, N.~H.~Hamann\cite{niko},
R.~T.~Jones\cite{jones}, B.~Mou\"{e}llic, S.~Ohlsson, J.-M.~Perreau}
\address{CERN, Geneva, Switzerland}

\author{W. Eyrich, M.~Moosburger, S.~Pomp, F.~Stinzing}
\address{University of Erlangen-N\"{u}rnberg, Erlangen, Germany}

\author{H.~Fischer, J.~Franz,  E.~R\"{o}ssle, H.~Schmitt, H.~Wirth}
\address{University of Freiburg, Freiburg, Germany}

\author{A.~Buzzo,  K.~Kirsebom, M.~Lo Vetere,
M.~Macr\`{\i}, M.~Marinelli, S.~Passaggio, M.G.~Pia, A.~Pozzo, E.~Robutti,
A.~Santroni}
\address{University of Genova and INFN, Genova, Italy}

\author{P.~T.~Debevec, R.~A.~Eisenstein, P.G.~Harris, D.W.~Hertzog,
S.~A.~Hughes, P.~E.~Reimer\cite{paul}, J.~Ritter}
\address{University of Illinois, Urbana, Illinois, USA}

\author{R.~Geyer, K.~Kilian, W.~Oelert, K.~R\"{o}hrich, M.~Rook, 
O.~Steinkamp}
\address{Institut f\"{u}r Kernphysik, Forschungszentrum J\"{u}lich,
J\"{u}lich, Germany}

\author{H.~Korsmo, B.~Stugu}
\address{University of Oslo, Oslo, Norway}

\author{T.~Johansson}
\address{Uppsala University, Uppsala, Sweden}
 \date{June 26, 1997}
\maketitle

\begin{abstract}

The $\fpptoksks$ cross section was measured at incident anti\-proton
momenta between 0.6 and 1.9~GeV/$c$ using the CERN Low Energy
Antiproton Ring (LEAR).  This investigation was part of a systematic
study of in-flight antiproton-proton annihilations into
two-neutral-meson final states in a search for hadronic resonances.  A
coarse scan of the $\pptoksks$ cross section as a function of
center-of-mass energy between 1.964 and 2.395~GeV/$c^{2}$ and a fine
scan of the region surrounding the $\xt$ are presented.  Upper limits
on the product branching ratio $\brxtopp\times\brxtoksks$ are
determined for a wide range of mass and width assumptions based on the
non-observation of the $\xt$\@. A rise in the $\pptoksks$ cross
section is observed near 2.15~GeV/$c^{2}$, which is consistent with
the $f_{2}(2150)$ resonance.
\end{abstract}

\section{Introduction}

Quantum Chromodynamics (QCD) has been very successful in describing
the strong interaction at high energies.  Within the framework of QCD,
hadrons are composed of colored quarks ($q$), antiquarks ($\qbar$) and
gluons ($g$) bound together into color neutral states.  The
experimentally observed families of bound states can be grouped and
described in the framework of the naive quark model in which only
three-quark ($qqq$) and quark-antiquark ($\qqbar$) constructions are
used.  
The fact that gluons as well as quarks carry color charge in QCD means 
that they should appear along with quarks as valence particles in 
hadronic wavefunctions. 
QCD calculations on the
lattice support the existence of states with valence glue 
and predict their masses with increasing
reliability~\cite{lattice}.  The experimental discovery of the
glueball spectrum would greatly increase our understanding of the
strong force at the hadronic scale~\cite{qcdreview}.

While no gluonic state has been conclusively identified, several
strong candidates exist.  Good arguments have been made that one or
both of the $f_{0}(1500)$ and the $f_{J}(1700)$ states might be a
scalar gluonic state or at least mixed with such a
state~\cite{amsler-1}.  At higher mass, the flavor-neutral decay
pattern and narrow width of the $f_{J}(2220)$, also known as the
$\xt$, have led to its identification as a possible tensor
glueball~\cite{huang}.  Additionally, arguments have been made in
support of the gluonic nature of the three broad tensor ``$g_{T}$
states'' at masses of 2.010, 2.300, and 2.340~GeV~\cite{etkin-1}.

The Jetset experiment was designed to search for such states by
measuring the energy dependence of the total and differential cross
sections of proton-antiproton annihilations into exclusive two-meson
final states.  The $\phi\phi$, $\ksks$ and $\eta\eta$ final states
were emphasized because of their suggested sensitivity to specific
candidate resonances and their expected small non-resonant cross
sections.  Incident antiproton momenta from 0.6 to 1.9~GeV$/c$ (1.964
to 2.395~GeV/$c^{2}$ in center-of-mass energy) were used.  The choices
of momenta provided both a broad scan of the entire energy region
available at LEAR, as well as a more focused study in the vicinity of the
$\xt$ state.

The $\xt$ was first reported by the Mark~III collaboration in
radiative $\jy$ decays~\cite{markiii}.  It appeared as a very narrow structure with a
mass of 2.231~GeV/$c^{2}$ and a width of 0.020~GeV/$c^{2}$ in the
reconstructed mass spectra of $\kp\km$ and $\ksks$ from the decays
$\jy\rightarrow\gamma\kp\km$ and $\jy\rightarrow\gamma\ks\ks$.  The
quantum numbers allowed for this state are $J^{PC} = ({\rm
even})^{++}$.  More recently, the BES experiment at
Beijing reported~\cite{bes} seeing the $\xt$ not only in $\kkbar$ decays but
also in non-strange $\jy\rightarrow\gamma\pi\pi$ and
$\jy\rightarrow\gamma\pbarp$ channels.  In this context a measurement
of the $\pptoksks$ cross section in the region of the $\xt$ is of
particular interest since both entrance and exit channels have been
observed to couple to this state.  A similar measurement
has also been reported by the PS185 collaboration at LEAR
\cite{ps185}.  In combination, these two experiments place strict
limits on the production of the $\xt$ in this channel.  Additional
motivation for the experiment is drawn from the reported
resonances in the $\phi\phi$ system at Brookhaven~\cite{etkin-1} 
in the reaction
$\pi^{-}p \rightarrow n\phi\phi$ and from the fact that data on
the meson spectrum from $\pbarp$ in-flight annihilations are 
comparatively scarce.

\section{Experiment}
\begin{figure}[tb]
  \begin{center}
    \mbox{\epsffile{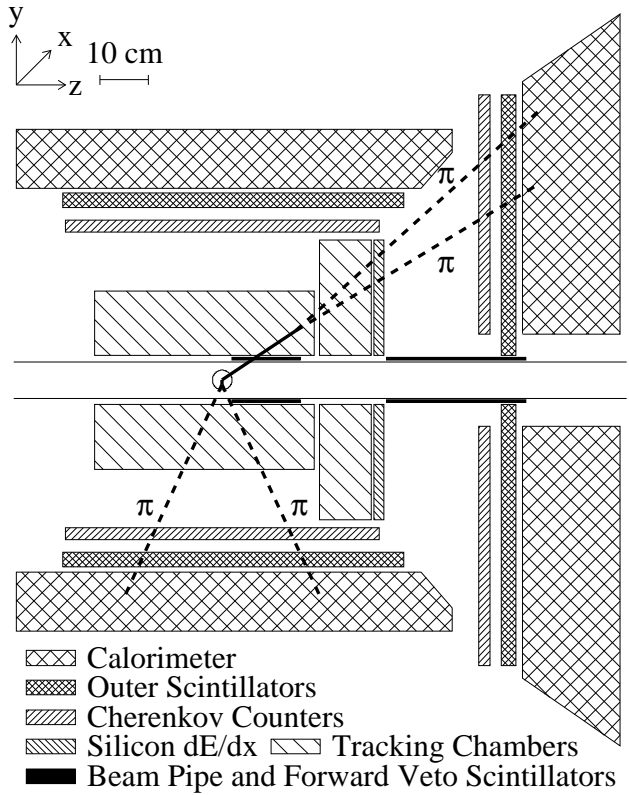}}
  \end{center}
    \caption{Schematic view of the Jetset detector with a $\ksks$
             event superimposed.}
  \label{detect}
\end{figure}

Data were collected using a non-magnetic detector constructed around a
hydrogen gas jet target installed in one of the straight sections of
the CERN Low-Energy Anti\-proton Ring (LEAR).  The detector was
divided into a forward end-cap covering the region from 9\degree\ to
45\degree\ and a barrel sector covering 45\degree\ to 135\degree\@.
Each region consisted of the following components: inner trigger
scintillators, straw tracking chambers, silicon dE/dx pads, 
threshold \v{C}erenkov counters, three layers of outer
scintillators, and an electromagnetic calorimeter.  A schematic view
of the detector is shown in Fig.~\ref{detect}.  More detailed
descriptions of the detector may be found elsewhere
\cite{evangelista}\cite{perthesis}.  The hydrogen gas jet target had a
density of up to $5\times 10^{12} \rm{~atoms/cm}^{2}$ at the beam
intersection.  LEAR typically stored between $2.5$ and $3.0\times
10^{10}$ antiprotons.  With a revolution frequency of approximately
3.2 MHz at a momentum of 1.5 GeV/$c$, this leads to an instantaneous
luminosity of approximately $4\times 10^{29} \rm{cm}^{-2}\rm{s}^{-1}$.
The fractional momentum uncertainty was less than 0.1\% or
approximately 0.5~MeV in center-of-mass energy.

The prominent feature of $\fpptoksks$ events is the relatively long
lived $\ks$ which has a mean lifetime, $\tau$, of 0.08926~ns, or
$c\tau = 2.676$~cm \cite{pdb}.  This allowed the $\ks$ mesons to
travel a macroscopic distance before decaying.  This feature was
exploited to detect and identify $\ksks$ events.  The tracks made by
the charged pions from the $\ks$ decay formed an unmistakable $V^{0}$
pattern in the detector.  The online identification of $\ks$ event
candidates looked for these delayed decays by requiring signals in the
outer scintillators and the \v{C}erenkov counters for each of the
charged pions while using the inner scintillators which surrounded the
interaction region as a veto shield.  This ensured that at least the
forward-going pair of pions was produced outside of the target region.

Data with the $\ks$ trigger were collected in July and October of
1991.  The July data set consisted of eight evenly spaced momenta
between 1.2 and 1.9 GeV/$c$, referred to as the ``coarse scan.''  The
October ``fine scan'' data included seven momenta from 1.39 to
1.48~GeV/$c$\/ in 0.015~GeV/$c$\/ steps, covering the $\xt$ mass
region.  In addition, $\ks$ data at 0.61 and 0.85 GeV/$c$\/ were
obtained during special calibration runs at the start of the October
period.  

\section{Data Analysis}

The reconstruction of events depended on the identification of a
delayed $\ks$ decay vertex.  Charged particle tracks were
reconstructed based on information from the straw tracking chambers.  To
form two independent vertices, four tracks were required.  An event 
sample with four or five tracks was examined for vertex combinations.
Vertices
were divided into two categories: those made of two forward tracks,
and those made of two barrel tracks\footnote{Vertices formed from one 
forward and one barrel track were not used.  The additional acceptance (less than 
$10\%$) owing to this topology was overwhelmed by the increase in background.  No clean 
extraction of such events was possible.}. 
In each event, forward and barrel vertex candidates were
made by matching all pairs of forward or barrel tracks.  Candidate vertices 
were removed if the distance of closest
approach of the two tracks was greater than $3.06~$cm or $5.89~$cm
for forward and barrel vertices, respectively, or if the
plane defined by the two tracks did not contain the target.

The barrel tracker straw wires provided modest (1.3 cm) 
position resolution in
the $z$ direction through charge division and good ($150 - 500~\mu$m) 
resolution in the
orthogonal coordinates ($xy$) from the drift-time information.  The latter 
fact was exploited to make a test of momentum
conservation in the $xy$ plane of the barrel $\ks$ decay vertex by
requiring that it is possible to draw a line from the interaction
region to the vertex, which extends through the opening of the $V^{0}$.
Barrel vertices failing this test were discarded.  Once identified,
each candidate vertex was geometrically fit.  The tracks from the fit
vertex were followed outward.  If either track passed through an inner
scintillator, which should have vetoed the event, the vertex was
removed.  This eliminated vertices made by random coincidences of tracks
which at some point along their trajectories passed the distance of
closest approach and other cuts mentioned earlier.  Events
with less than two independent vertices and events containing photons,
identified by the calorimeter, were removed from the event sample.

The momentum of each tracked pion was not directly measured, but was
determined through solution of momentum and energy
conservation in the event.  This solution assumed that the
reconstructed tracks were produced by pions and was based on the
measured directions of the particles.  For each event, up to two
solutions could be consistent with the kinematics.  
Owing to the finite detector resolution, a ``violation'' of energy
conservation for a candidate solution was permitted up to
$0.2 \times E_{beam}$.
Monte Carlo studies verified the placement of this cut.  
For events with no solutions, either the
hypothesis that the tracks were produced by pions was wrong, or there
were other, unobserved final state particles in the event.  In either
case, these events were removed from the event sample.

\begin{figure}
  \begin{center}
   \mbox{\epsffile{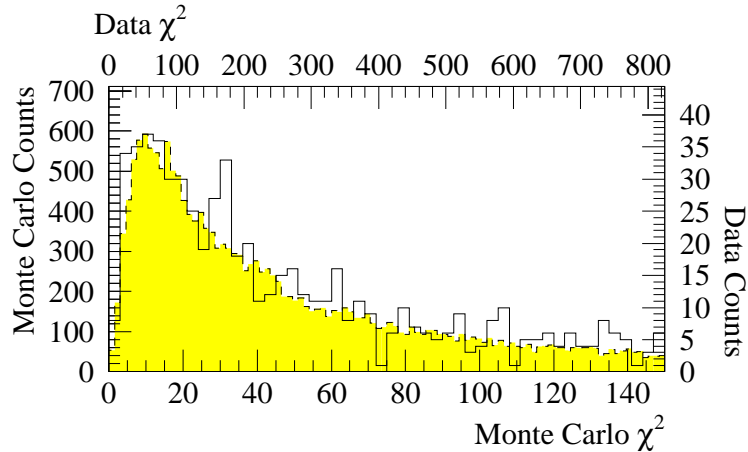}}
  \end{center}
     \caption{$\chi^{2}$
           distributions for data (unshaded, solid line) and Monte
           Carlo events (shaded, dashed line).  The lower and left
           scales are for the Monte Carlo events and the upper and
           right scales are for data.}
  \label{fig:chi}
\end{figure}

A least squares fit to the kinematics of $\fpptoksks$ was
performed for each event having an allowable momentum solution.  The
kinematic fit provided improved precision for the momenta of the pion
tracks, and yielded a measure of the probability that the event matched
the $\fpptoksks$ hypothesis.  The $\chi^{2}$ distribution is shown in
Fig.~\ref{fig:chi}.  For both data and Monte Carlo events, the
$\chi^{2}$ distribution was found to be broader than that 
expected for an {\em ideal\/} least squares fit with six degrees of
freedom.  This was not surprising owing to effects such as
multiple-scattering and pion interactions.  When compared in detail, the
Monte Carlo and data $\chi^{2}$ distributions have a nearly identical
shape, however with the scale stretched by a factor of 5.5
for the real events.  This factor comes from consideration of the
additional non-Gaussian uncertainties which are present in
the detector but which were not included in the simulation.  For
example, the precision of the positioning of the forward and barrel
trackers with respect to one another was worse than the
resolution of these devices.  Further, the true straw-tracker
resolution function had to be described by two Gaussians, one narrow
and one broad.  In the simulation, only an average was used.  Both
effects were studied and were found to contribute to the
$\chi^{2}$ scale difference.  The maximum allowed $\chi^{2}$ in the
final data sample was set to 825 which was large enough so that the scale
difference did not affect any of our conclusions beyond the systematic
errors we report.
 
These steps led to the following reduction in the number of events.  The
fraction of two-vertex candidate events in the raw trigger was $0.045\%$
and $0.079\%$ for the fine- and coarse-scan data sets, respectively.  This
difference is understood and is described later. Of the 6195 fine-scan and
11,442 coarse-scan events which remained, approximately $16\%$ were left
once the photon cut was applied.  Events with candidate kinematical
solutions lowered the sample to approximately $9\%$.  Finally, after the
kinematic fitting and $\chi^{2}$ cuts were applied 159 fine-scan and 346
coarse-scan events remained.  We estimate that approximately 90\% of these
events were $\pptoksks$ events as discussed below. 

\begin{figure}
  \begin{center}
   \mbox{\epsffile{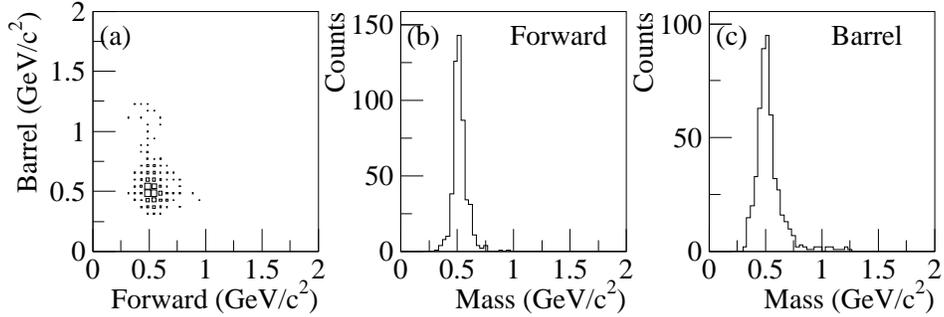}}
  \end{center}
     \caption{Goldhaber plot of invariant mass combinations formed from
    forward and barrel vertices.  This plot represents all of the data
    where a solution has been found with two independent vertices.}
  \label{fig:gh}
\end{figure}

A plot of the invariant mass for the forward vertex versus that of the
barrel\footnote{The kinematics of $\pptoksks$ forbids
events with both vertices in the barrel region or both vertices in
the forward region, except in rare instances at the 1.9~GeV/$c$ incident
momentum setting.} is shown in Fig.~\ref{fig:gh} for events which, when
processed by the event fitting routine, were found to have an 
acceptable $\chi^{2}$.  The
plot is dominated by events in the $\ksks$ mass region.  Evidence for
a small background contamination can be seen in the regions on the
high-mass sides of the peak.  The background events are primarily from
the $\spptokskr$ and $\pptokskrc$ reactions and their non-resonant
partners, which can easily mimic the channel of interest.  These
reactions have cross sections from 10 to 100 times larger than the
$\pptoksks$ cross section~\cite{flaminio}.  The final states involving
neutral $\piz$ are identified and rejected when one or more photons
were observed in the calorimeter.  Monte Carlo studies of simulated
background events processed according to the $\ksks$ hypothesis
confirmed that a small number do enter the final sample and that the
invariant mass reconstruction for these events is always on the high
side of the $\ks$ mass as seen in the plot.

The number of background events was estimated by comparing the decay
length distribution with the ideal one for true $\ks$ decays.  The
measured decay lengths, converted to a lifetime distribution, should
show an exponential decay.  Any deviation from this due to events with
charged particles emerging directly from the origin shows up as an
excess at very small lifetimes, while the tail of the distribution is
unaffected.  The background was estimated by fitting the tail of the
lifetime distribution and extrapolating into the region where the
prompt $V^{0}$ decays create an excess.  The exact shape of the distribution
for the delayed decays was determined by Monte Carlo taking into
account the full acceptance of the detector.

\begin{figure}
  \begin{center}
    \mbox{\epsffile{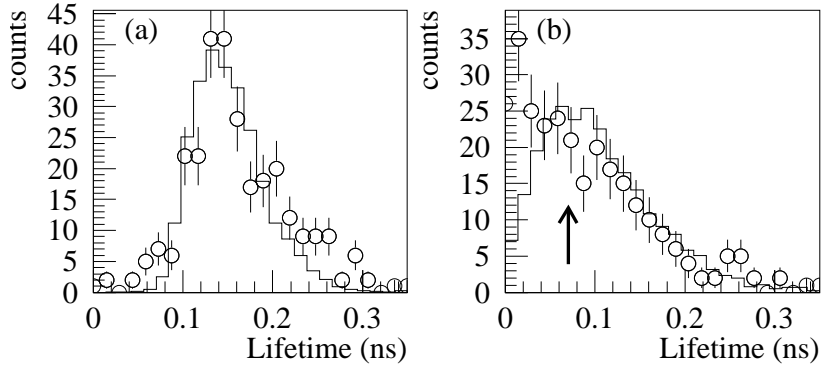}}
  \end{center}
  \caption{Lifetime distributions for the fine-scan data
           (points) and Monte Carlo (histogram) for forward (a) and
           barrel (b) vertices.  The arrow indicates where the
           ``tail'' began for purposes of the fit.}
  \label{lifesum}
 \end{figure}

This method of estimating the number of background events in the
sample depends on having a statistically significant number of events
in the tail of the lifetime distribution.  For this purpose, the data
were divided into three groups: the 1.6 to 1.9~GeV/$c$ coarse-scan data, the
1.2 to 1.5~GeV/$c$ coarse-scan data and the fine scan (1.39 to
1.50~GeV/$c$).  Even with this division, the statistics in the tail
region were limited, and a ``binned likelihood'' procedure based on
Poisson statistics was used for the fit~\cite{binfit}.  The only
parameter varied in this fit was an overall scale factor for the Monte
Carlo distributions.  Based on this procedure, the event samples
contained $(89.1\pm9.3)\%$, $(91.8\pm7.2)\%$ and $(91.5\pm8.9)\%$ true
$\pptoksks$ events for the three samples, respectively.  Fig.~\ref{lifesum} shows the
lifetime distributions for the fine-scan data along with a
corresponding Monte Carlo distribution which has been scaled by the
fitting procedure described.

The Monte Carlo events were generated using the GEANT package from
CERN~\cite{geant}.  The geometry and composition of all detectors and
relevant support structures were included in the simulation.  Monte
Carlo events for the reaction $\pptoksks$ were generated isotropically
in $\cos(\theta_{cm})$ and then weighted according to the second order
Legendre polynomial fit reported by the PS185 collaboration for their
data obtained near 1.43~GeV/$c$ incident antiproton 
momentum~\cite{ps185}.  Since the true $\ksks$
differential cross section rises slightly at forward angles, this
procedure resulted in an increase in the inferred total cross section
by an average of 4\% compared to what would have been obtained
assuming a flat differential cross section.  The Monte Carlo events
passed through the same analysis steps outlined above to determine the
overall detector acceptance.  Events from relevant background
reactions such as $\pptokskr$ were also generated with a uniform
angular distribution and were studied for feed-down into the
$\ksks$ sample.

The integrated luminosity at each momentum setting was determined by
continuously measuring the $\pbarp$ elastic differential cross
section at 90\degree\ in the center-of-mass frame.  The absolute cross section
for elastic scattering is well established throughout the energy region of 
interest~\cite{ppelastic}.  A special trigger, based on pairs of pixels 
in the forward outer scintillator array, was used to select these
events and a separate analysis was performed on this data sample.  A
comparison of the acceptance-corrected elastic yield to
the published cross section provided a measure of the absolute
integrated luminosity.  An additional 15\% uncertainty, not included in the 
errors on the individual points, exists on the scale of our final cross
sections and is common for all of the energy points.  This systematic
error includes 5\% uncertainty on the $\pbarp$ elastic cross
sections.  The relative energy-point to energy-point integrated luminosity is
more precise and was derived not only from the elastic event sample,
but also from a combination of trigger scintillator scalers which were
found to be very reliable and stable for the lifetime of the Jetset
experiment.  The relative luminosity error was found to be approximately
$2\%$.

\section{Results and Discussion}

\begin{figure}
  \begin{center}
    \mbox{\epsffile{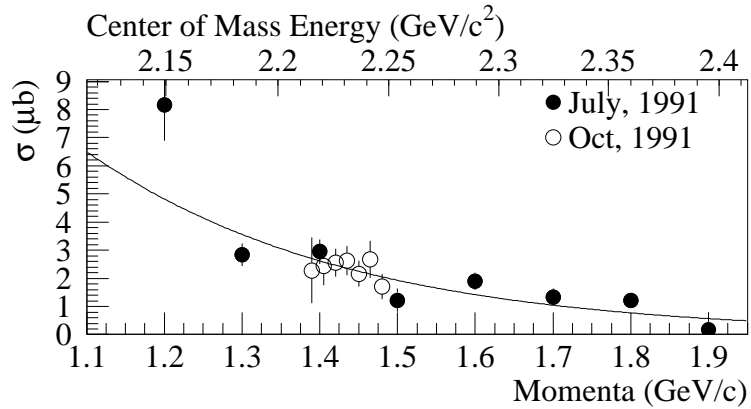}}
  \end{center}
     \caption{Total cross section for the reaction $\pptoksks$.  The
    solid curve is a fit following the form $Ap^{*} \times e^{-B
    p^{*}}$, excluding the 1.2 GeV/$c$ point.  The fine-scan data have
    been adjusted to match the scale of the coarse-scan by means of a
    multiplicative factor. }
  \label{totalxs}
\end{figure}

The total cross section as a function of the center-of-mass energy was
derived by dividing the background-subtracted number of observed
events by the integrated luminosity and by the acceptance.  The results are
listed in Table~\ref{totaltab} and are shown in
Fig.~\ref{totalxs}. The measured cross section is interpreted as the
sum of contributions from a smoothly-varying, non-resonant production
plus any resonances.  The exact form of the non-resonant component of
the cross section, $\sigma_{nr}$, is not important as it changes only
slightly over the region of interest and several functional forms can
be used.  However, we find it convenient to employ the
parametrization of Vandermeulen \cite{vandermeulen}, which recognizes
that $\Npair$ annihilations proceed dominantly through two-meson
intermediate states.  It has the form
\begin{equation}
\sigma_{nr} = Ap^{*} \times e^{-B p^{*}},
\label{eq:sig_nr}
\end{equation}
where $A$ and $B$ are parameters, and $p^{*}$ is the momentum of the
$\ks$ mesons in the center-of-mass,
$p^{*}=(1/2)\times\sqrt{s-4m_{\ks}}$.  The solid line in
Fig.~\ref{totalxs} follows this form and fits the data well.

\begin{table}
\begin{center}
\begin{tabular}{|c|c|c|c|c|c|}
\hline
Beam     & Center  &  Integrated & Monte   & Num.   & Cross \\
Momentum & of Mass &  Luminosity & Carlo   & of     & Section\\ 
         & Energy  &             & Accept. & Events & ($\mu$b)\\ 
(GeV/$c$)&(GeV/$c^{2}$)
                   & (nb$^{-1}$) &  (\%)   &        & \\ \hline 
\multicolumn{6}{|c|}{Coarse Scan} \\ \hline  
1.900 & 2.395 &  3.59 & 0.31 &  1 & $ 0.17\pm 0.17$ \\
1.800 & 2.360 & 20.75 & 0.31 & 41 & $ 1.21\pm 0.23$ \\
1.700 & 2.324 & 11.17 & 0.32 & 25 & $ 1.33\pm 0.30$ \\
1.600 & 2.289 & 19.48 & 0.33 & 65 & $ 1.90\pm 0.31$ \\
1.500 & 2.254 &  4.13 & 0.35 &  9 & $ 1.22\pm 0.42$ \\
1.400 & 2.218 & 13.71 & 0.33 & 69 & $ 2.95\pm 0.43$ \\
1.300 & 2.183 & 16.10 & 0.33 & 77 & $ 2.84\pm 0.40$ \\
1.200 & 2.149 &  4.75 & 0.30 & 59 & $ 8.16\pm 1.26$ \\\hline
\multicolumn{6}{c}{} \\ \hline
\multicolumn{6}{|c|}{Fine Scan} \\ \hline 
1.480 & 2.247 &  8.62 & 0.085 & 17 & $ 1.71\pm 0.45$ \\
1.465 & 2.241 &  6.06 & 0.092 & 20 & $ 2.67\pm 0.66$ \\
1.450 & 2.236 & 11.29 & 0.085 & 29 & $ 2.17\pm 0.46$ \\
1.435 & 2.231 & 11.57 & 0.085 & 36 & $ 2.63\pm 0.52$ \\
1.420 & 2.225 & 11.92 & 0.092 & 38 & $ 2.56\pm 0.49$ \\
1.405 & 2.220 &  5.30 & 0.085 & 15 & $ 2.44\pm 0.68$ \\
1.390 & 2.215 &  1.49 & 0.085 &  4 & $ 2.28\pm 1.17$ \\\hline
\multicolumn{6}{c}{} \\ \hline
\multicolumn{6}{|c|}{Additional Points} \\ \hline
0.850 & 2.033 &  1.76 & 0.037 &  7 & $ 6.04\pm 2.76$ \\
0.609 & 1.964 &  1.03 & 0.018 &  2 & $ 6.07\pm 5.06$ \\
\hline
    \end{tabular}
  \end{center}

  \caption{The $\pptoksks$ cross sections.  Also listed are the
           integrated luminosity, the acceptance, the number of events
           detected at each energy and the fraction of those events
           which are $\fpptoksks$ events.}

  \label{totaltab}
\end{table}

Between the coarse-scan and fine-scan runs, the radiator in the \v{C}erenkov
counters was changed from FC72 \cite{fc72} to water with a
corresponding decrease in the threshold $\beta$ 
from $0.79 \times c$ to $0.75 \times c$.  
This detector was designed to be used as a veto for fast pions 
($\beta > 0.9$) in a trigger for the reaction $\pptofourk$ which was a primary 
channel of our experiment. To keep the fraction of false fast-pion vetos low,  
the discriminator thresholds on these detectors were raised so that $25 - 
50\%$ of the fast charged pions passed through undetected.  This 
change in threshold happened when between the coarse- and fine-scan 
runs.
The consequence  
to the $\ksks$ data set was  
an inefficiency in the trigger for the fine scan.  Since the thresholds were 
comfortably
below pion threshold during the coarse scan, we choose to normalize 
the fine-scan data to the coarse-scan data to establish the final 
cross section values.   The
ratio $A_{coarse}/A_{fine} = 0.61\pm0.07$ was used.  In the search for 
narrow resonances in the fine-scan data, this normalization factor 
does not influence the relative cross section data, only the scale.

\begin{figure}
  \begin{center}
    \mbox{\epsffile{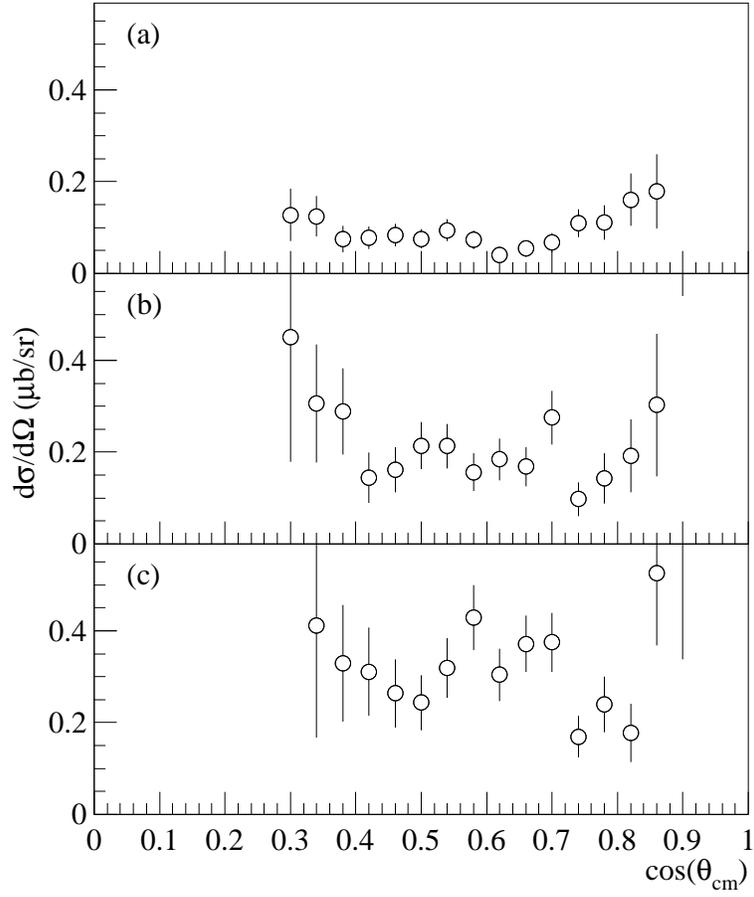}}
  \end{center}
     \caption{Differential cross section results for three energy 
    groupings of the reaction $\pptoksks$.  The 1.2, 1.3, 1.4 and 
    1.5~GeV/$c$ data are shown in (a), the fine scan 1.39-1.48~GeV/$c$ 
    data are shown in (b) and the 1.6, 1.7, 1.8 and 1.9 GeV/$c$ data 
    are shown in (c).}
  \label{diffxs}
\end{figure}

To calculate the differential cross section with sufficient
statistics, the data were summed into the three groups mentioned above. 
These distributions are shown in
Fig.~\ref{diffxs}.  The angular acceptance of the detector limited
the range of the differential cross section to $0.3 < cos(\theta_{cm})
< 0.9$.  In the region of the acceptance, all of the distributions 
show little structure.  A Legendre polynomial fit was not done due to the
limited data near $cos(\theta_{cm}) = 1$, which is needed in order to
include the higher-order terms.

The resonant portion of the cross section, if present, may be
described by a Breit-Wigner line shape.  If there is no interference between
the resonance and the background, then these contributions can be
summed.  The Breit-Wigner parametrization is given by
\begin{equation}
\sigma_{BW} = (w_{i}w_{f}) \times
              \frac{(2J+1)}{(2S_{1}+1)(2S_{2}+1)} \times
              \frac{4\pi(\hbar c)^{2}}{s-4m_{p}^{2}} \times
              \frac{\Gamma^{2}}{(\sqrt{s}-m_{res})^{2} + \Gamma^{2}/4}.
  \label{eq:bw}
\end{equation}
Here $(w_{i}w_{f})$ is the double branching ratio and $w_{i}w_{f} =
\rm{BR}(X\rightarrow\pbarp)\times\rm{BR}(X\rightarrow\ks\ks)$\@. The
$S_{i}$ terms are the spins of the initial proton and antiproton
(1/2), and $J$ is the total angular momentum of the resonance,
reducing the angular momentum term, $(2J+1)/ \left(
\left(2S_{1}+1)(2S_{2}+1\right)\right)$ to either $5/4$ or $9/4$ for 
$J = 2$ or $J = 4$.  With these parametrizations of the
non-resonant and resonant cross sections, the total cross section may
be expressed as a function of five parameters: $A$ and $B$ from
Eq.~\ref{eq:sig_nr} and $(w_{i}w_{f})$, $\Gamma$ and $m_{res}$ from
Eq.~\ref{eq:bw}.  To completely describe the data, a sixth parameter
was added to renormalize the fine scan to the coarse-scan data.

For fixed mass and width assumptions, the likelihood ratio test was
used to place limits on the strength of a possible resonance
\cite{frodesen}.  In this test, an initial likelihood fit was made in
which the strength of the resonance was allowed to vary freely.  The
fit was then repeated with a fixed resonance strength,
$(w_{i}w_{f})^{*}$.  The likelihood ratio is defined as $\lambda =
\frac{L^*}{L}$, where the likelihood from the initial fit is $L$ and
the likelihood from the fit with fixed resonance strength is $L^*$.  The
significance of the resonance strength, $(w_{i}w_{f})^*$, can be
deduced by noting that the statistic $-2\ln{\lambda}$ follows a
$\chi^2$ distribution with one degree of freedom.  The resonance
strength was systematically increased and new fits were made until the
resonance strength corresponding to a significance 0.05 was found.
This strength represents an upper limit on the double branching ratio
with confidence of 95\% for the particular mass and width which were
chosen.

\begin{figure}
  \begin{center}
    \mbox{\epsffile{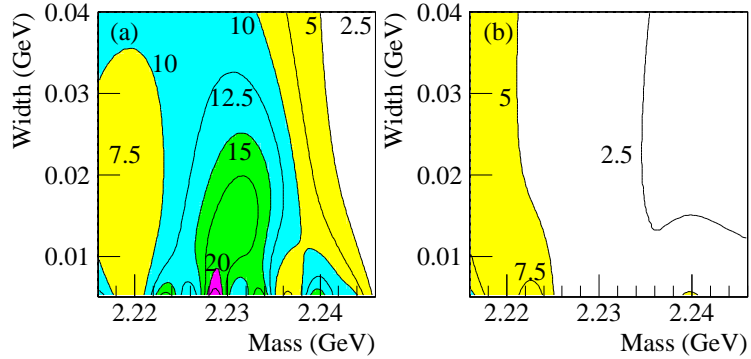}}
  \end{center}
     \caption{Contours of the upper limit on the double branching ratio
             $\brxtopp\times\brxtoksks$ as a function of resonance
             mass and width for a $J = 2$ resonance.  The contours
             represent steps of $2.5\times 10^{-5}$ and the shading
             changes in steps of $5\times10^{-5}$.  In (a) only the
             data from this work are considered.  In (b) our data are
             combined with the data from PS185~\protect\cite{ps185}
             and the same procedure is repeated.}
  \label{contours}
\end{figure}

In the region of the $\xi$, this procedure was performed for widths in
the range from 5 to 40~MeV/$c^{2}$ and masses from 2.219 to
2.246~GeV/$c^{2}$.  The composite results are compiled in
Fig.~\ref{contours}a in the form of a contour map representing upper
limits on the double branching ratio as a function of resonance width
and mass.  A tensor ($J = 2$) resonance was assumed.  For a $J = 4$
resonance, the upper limit must be multiplied by the factor $5/9$ to
account for the spin term in Eq.~\ref{eq:bw}.  The least restrictive
limit on a possible resonance as determined by this data set alone 
occurs at a mass of 2.231~GeV and a width of 0.012~GeV.  Here, the
upper limit on the double branching ratio is $19.5 \times 10^{-5}$.
A fit of the cross section which forces a resonance at this
point produces a double branching ratio of approximately $5 \times
10^{-5}$ with a significance of just less than one standard deviation.

A similar analysis was made for the cross section measurements
presented here combined with those reported by the PS185
collaboration~\cite{ps185} which cover the same general energy region,
however at slightly different specific momentum values.  Merging the
two data sets required an additional parameter to account for the 15\%
uncertainty in the global normalizations reported as systematic errors
in the overall cross section scale by each experiment.  The contours
for the combined data are shown in Fig.~\ref{contours}b.  Some
insensitive regions exist where neither experiment accumulated data,
however generally the limits\footnote{The $3\sigma$ limits quoted in 
Ref.~\protect\cite{ps185} are more restrictive than those produced 
by the likelihood method when applied to either the
PS185 data alone, or to the combined PS185/Jetset data.   
A constrained background function was used in the PS185 fitting 
procedure which 
may contribute in part to the difference.  We judge the likelihood method and 
results to be more general in nature and properly representative 
of the true limits implied by these searches for the $\xi$.} 
on the double branching ratio are
relatively constant at less than $7.5 \times10^{-5}$ for a resonance
whose width is greater than 5~MeV/$c^{2}$.

\begin{figure}
  \begin{center}
    \mbox{\epsffile{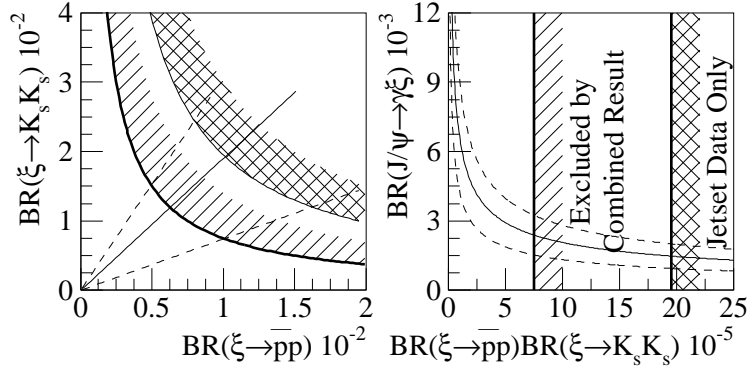}}
  \end{center}
     \caption{Regions of allowed branching ratios.  The solid lines
    represent the observed strengths of coupling of the $\xi$ to
    $\ksks$ (MARK III and BES) and $\pbarp$ (BES only).  The dashed
    lines are $1 \sigma$ limits.  A portion of each plot is excluded 
    by use of the Jetset data alone (cross hatched).  Significantly 
    tighter limits are set by the combined Jetset/PS185 data (single 
    hatched).} 
  \label{limits}
\end{figure}

These results may be combined with those from Mark~III and BES to
establish allowed values for the single branching ratios $\brjtogxi$,
$\brxtoksks$ and $\brxtopp$\@.  Plotted in Fig.~\ref{limits}a are the
single branching ratios $\brxtoksks$ versus $\brxtopp$\@.  The $\ksks$
are averaged from MARK III and BES, while the $\pbarp$ result is from
BES alone\footnote{ MARK III set a limit of $\brjtogxi\times\brxtopbarp 
< 2\times10^{-5}$ at a confidence limit of 90\%.  BES measured
$1.5\times10^{-5}$ for the same quantity based on a peak with a 3.8
standard deviation significance.}.  The results presented here form a
hyperbola which sets an upper bound on the product, implying that
single branching ratios up to the 1 to $2\%$ limit are allowed.  Very
large couplings are excluded.  In Fig.~\ref{limits}b, the same
information is used to show the region permitted for the coupling
$\jy\rightarrow\gamma\xi$.  In the allowed region, the branching ratio
is greater than $0.2\%$ which indicates a very strong coupling.  An
upper limit on $\brjtogxi$ may also be inferred based on the total of
all radiative decays which is approximately $8\%$.

An examination of the broad scan reveals that the $\pptoksks$ cross
section at 1.2~GeV/$c$ (2.148 GeV/$c^2$) appears significantly larger
than is expected with the simple non-resonant parametrization.
During the period of the fine scan, data were collected at two 
additional momenta below
1.2~GeV/$c$.  These data, 0.61 and 0.85~GeV/$c$, were used for
detector calibration.  The $\pptoksks$ trigger was in operation, but
the luminosity trigger was not.  The integrated luminosity for these
points was deduced by comparing the counting rates in simple
coincidences formed by various triggering counters with those obtained
at other momenta where the integrated luminosity was known.  An extrapolation was
used for the anticipated momentum-dependence of the rates.  The final
values for these low-momentum points are given in Tab.~\ref{totaltab}
where the errors not only reflect the small number of events, but also
the additional uncertainties in the integrated luminosity and acceptance.

\begin{figure}
  \begin{center}
    \mbox{\epsffile{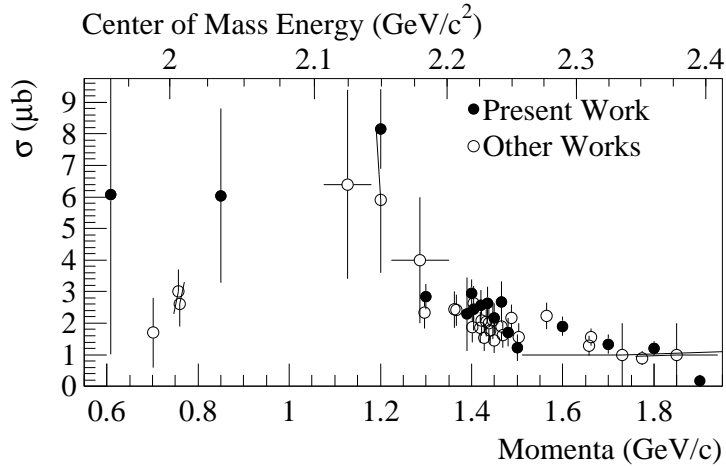}}
  \end{center}
     \caption{The world's data for the reaction $\pptoksks$.  Other
    data are taken from Refs.~[8,12,19-27].}
  \label{world}
\end{figure}

A complete view of the $\pptoksks$ total cross section is seen in
Fig.~\ref{world} where all the world's data are represented.  A
resonance in the region of the 1.2~GeV/$c$ point has been reported in
several other channels including $\pptopizpiz$~\cite{dulude}, 
$\pbarp\rightarrow\pbarp$~\cite{coupland}, $\pptonn$~\cite{cutts}, 
and the $\pbarp$
total cross section~\cite{alspector}.  All find evidence for a
structure near 2.150 GeV with a statistically consistent width in the
range from 0.050 to 0.250 GeV/$c^{2}$.  The fit by the BES
collaboration of the $\pbarp$ spectrum in $\jy\rightarrow\gamma\pbarp$
also included such a structure at 2.144~GeV/$c^{2}$~\cite{bes}.  A
$2^{++}$ resonance, known as the $f_{2}(2150)$, is associated with
this collection of observations by the Particle Data Group~\cite{pdb}.
When the data below $1.8~\rm{GeV}/c^{2}$ in the $\pptoksks$ summary
are fit with a freely floating Breit-Wigner permitted to sum
incoherently with the background, one finds that the data is
consistent with this resonance, having a double branching ratio of
$(w_{i}w_{f}) = (27^{+9}_{-7})\times 10^{-5}$ at
$2.139^{+0.008}_{-0.009}~\rm{GeV}/c^{2}$ having a width of
$0.056^{+0.031}_{-0.016}~\rm{GeV}/c^{2}$.  
The $\chi^2$ per degree of
freedom for this fit was 28.9/40 compared with 60.7/43 when fit with no
resonance.

\section{Conclusions}
We have reported new results on a search for the $\xt$ in the
formation channel $\pptoksks$.  No evidence for the $\xt$ was found.
Combining these results with those from earlier work at LEAR by the
PS185 experiment~\cite{ps185} sets new limits on the double branching
ratio product $\brxtopp\times\brxtoksks$ for a wide range of mass and
width assumptions of the $\xi$. For mass and width combinations
appropriate to the radiative $J/\psi$ findings, the double branching
ratios are all less than $7.5 \times 10^{-5}$ at a confidence level of
$95\%$.
The implications of this limit are that the coupling of the $\xt$ to
the final states $\kkbar$ and $\ppair$ is very small, at the level 
of 1\% or less.  Given these results in combination with other
measurements of the $\xt$, 
the question arises, if the channels in which the $\xt$ has been 
observed are not its primary decay modes, to which channels does it 
strongly couple?
In a study of the $\piz\piz$, $\eta\eta$ and $\phi\phi$ final states, 
our experiment establishes similar limits~\cite{etapaper}.
If the $\xt$ does indeed couple to $\pbarp$ at the level reported by 
BES, then greater than 90\% of its decays have yet to be discovered.

\section{Acknowledgments}
We thank the teams of the CERN Antiproton Complex, in particular the
LEAR staff. This work has been supported in part by CERN, the
German Bundesministerium f\"{u}r Bildung, Wissenschaft, Forschung und
Technologie, the Italian Istituto Nazionale di Fisica Nucleare, the
Swedish Natural Science Research Council, the Norwegian 
Research Council, and the United States National Science
Foundation, under contract NSF PHY 94-20787.  This work was based in
part on the dissertation of P.~E.~Reimer, submitted to the University
of Illinois in partial fulfillment of the requirements for the
Ph.D. degree.


\begin{references}
\bibitem[^{\dagger}]{niko} Deceased.
\bibitem[^{*}]{jones} Present Address: University of Connecticut, Storrs, CT.
\bibitem[^{**}]{paul} Present Address: Los Alamos National Laboratory, 
Los Alamos, NM. 

\bibitem{lattice}G. S. Bali \etal, Phys. Lett. B {\bf 309} (1993) 
378; and J. Sexton, A. Vacarino and D. Weingarten, Phys. Rev. 
Lett. {\bf 75} (1995) 4563. 

\bibitem{qcdreview}F.~E.~Close, Rep. Prog. Phys. {\bf 51} (1988)
833; and T.~H.~Burnett and S.~R.~Sharpe, Annu. Rev. Nucl. Part. Sci.
{\bf 40} (1990) 327.

\bibitem{amsler-1}C.~Amsler and F.~Close, Phys. Lett. {\bf B353}
(1995) 3858; C.~Amsler and F.~Close, Phys. Rev. {\bf D53} (1996) 295;
and J.~Sexton, A.~Vacarino and D.~Weingarten, Phys. Rev. Lett. {\bf
75} (1995) 4563.

\bibitem {huang} T. Huang, S. Jin, D. Zhang, K. Chao, Phys. Lett. 
{\bf B380} (1996) 189.

\bibitem{etkin-1}A.~Etkin \etal, Phys. Lett. {\bf B165} (1985) 217,
and A.~Etkin \etal, Phys. Lett. {\bf B201} (1988) 568.

\bibitem{markiii}R.~M.~Baltrusaitis \etal, Phys. Rev. Lett.  {\bf56}
(1986) 107.

\bibitem{bes}J.~Z.~Bai \etal, Phys. Rev. Lett. {\bf 76} (1996) 3502.

\bibitem{ps185}P.~D.~Barnes \etal, Phys. Lett. {\bf B309} (1993)
469.

\bibitem{evangelista}C.~Evangelista \etal, Phys. Lett {\bf B345}
(1995) 325; and C.~Evangelista \etal, {\em The $\pptophiphi$ reaction
near threshold,} CERN SPSLC 92-42, SPSLC M501 (10 August 1992).

\bibitem{perthesis}P.~E.~Reimer, Ph.D. Thesis, University of
Illinois, 1996.

\bibitem{pdb}Particle Data Group, Phys. Rev. {\bf D54}, (1996) 1.

\bibitem{flaminio}V.~Flaminio \etal, Compilation of Cross-Sections
III:  $p$ and $\pbar$ Induced Reactions,  CERN-HERA 84-01 (17 April
1984).

\bibitem{binfit}R.~Barlow and C.~Beeston, Comp. Phys. Comm. {\bf 77}
(1993) 219; and F.~James, Minuit: Function Minimization and Error
Analysis, CERN (1994).

\bibitem{geant} Application Software Group, Computing and Networks
Division, GEANT:  Detector Description and Simulation Tool, (CERN,
Geneva, Switzerland, 1993).

\bibitem{ppelastic}E.~Eisenhandler \etal, Nucl. Phys. {\bf B113}
(1976) 1-33; T.~Bacon \etal, Nucl. Phys. {\bf B32} (1971) 66-74; and
D.~L.~Parker \etal, Nucl. Phys. {\bf B32} (1971) 29-44.

\bibitem{vandermeulen}J.~Vandermeulen, Z. Phys. {\bf C37} (1988) 563.

\bibitem{fc72}FC72 is the trade name of a compound marked by 3M
composed of $C_{6}F_{14}$, sometimes known as ``liquid freon.''

\bibitem{frodesen} A.~G.~Frodesen, O.~Skjeggestad, and H.~T\o fte,
{\bf Probability and Statistics in Particle Physics}, Bergen:
Universitetsforlaget, 1979.

\bibitem{ganguli} S.~N.~Ganguli \etal, Nucl. Phys. {\bf B183} (1981)
295-329.

\bibitem{cooper}A.~M.~Cooper \etal, Nucl. Phys. {\bf B136} (1978)
365.

\bibitem{duboc}J.~Duboc \etal. Nucl. Phys. {\bf B46} (1972) 429.

\bibitem{barlow}J.~Barlow \etal, Nuovo Cimento {\bf A50} (1967) 701.

\bibitem{handler} T.~Handler \etal, Nucl. Phys. {\bf B110} (1976)
173-88.

\bibitem{oh}B.~Y.~Oh \etal, Nucl. Phys. {\bf B51} (1973) 57.

\bibitem{todenhagen} R.~Todenhagen, Ph. D. Thesis,
Albert-Ludwigs-University, Freiburg, Germany, 1995.

\bibitem{chapman}J.~W.~Chapman \etal, Nucl. Phys. {\bf B42} (1972) 1.

\bibitem{fields}T.~Fields \etal, Phys. Lett. {\bf B40} (1972) 503.

\bibitem{dulude}R. S. Dulude \etal, Phys. Lett. {\bf B79} (1978) 335.

\bibitem{coupland}M. Coupland \etal, Phys. Lett. {\bf B71} (1977) 460.

\bibitem{cutts}D. Cutts \etal, Phys. Rev. {\bf D17} (1978) 16.

\bibitem{alspector}J. Alspector \etal, Phys. Rev. Lett. {\bf 30} (1973) 511.

\bibitem{etapaper}A. Buzzo \etal, in preparation.

\end{references}
\end{document}